\documentclass[%
 amsmath,amssymb,
aps,pre,twocolumn,amsmath,amssymb,nofootinbib,superscriptaddress,floatfix
]{revtex4-2}

\usepackage{graphicx}
\usepackage{dcolumn}
\usepackage{bm}
\usepackage{physics}
\usepackage{xcolor}
\usepackage{bbm}

\newcommand{\xm}[1] {{\color{black}#1}}

\allowdisplaybreaks

\newcommand{\splitatcommas}[1]{%
\begingroup 
\begingroup\lccode`~=`, \lowercase{\endgroup 
\edef~{\mathchar\the\mathcode`, \penalty0 \noexpand\hspace{0pt plus 1em}}%
}\mathcode`,="8000 #1%
\endgroup 
} 

\begin{document}

\title{Stress control in non-ideal topological Maxwell lattices via geometry}

\author{Harry Liu}
\thanks{These  authors contributed equally.}
\affiliation{Department of Physics, University of Michigan, Ann Arbor, MI 48109, USA}

\author{Siddhartha Sarkar}
\thanks{These  authors contributed equally.}
 \affiliation{Department of Physics, University of Michigan, Ann Arbor, MI 48109, USA}

\author{A. Nafis Arafat}
\affiliation{Department of Physics, University of Michigan, Ann Arbor, MI 48109, USA}
\affiliation{City University of New York, New York, NY 10017, USA}

\author{Ethan Stanifer}
 \affiliation{Department of Physics, University of Michigan, Ann Arbor, MI 48109, USA}

\author{Stefano Gonella}
 \affiliation{Department of Civil, Environmental, and Geo- Engineering, University of Minnesota, Minneapolis, MN 55455, USA}

\author{Xiaoming Mao}
\email{maox@umich.edu}
 \affiliation{Department of Physics, University of Michigan, Ann Arbor, MI 48109, USA}

\date{\today}

\bibliographystyle{vancouver}

\begin{abstract}
Topological mechanical metamaterials have demonstrated exotic and robust mechanical properties which led to promising engineering applications. One of such   properties is the focusing of stress at the interface connecting domains of topological Maxwell lattices of opposite topological polarizations, which protects the bulk of the material against fracturing. Here we generalize this theory to non-ideal Maxwell lattices,  incorporate real material features that leads to interactions beyond previous ideal models.
By quantitative analysis of stress distributions of topological Maxwell lattices with self-stress interfaces theoretically and computationally, we propose a design rule that minimizes stress in the bulk of the material.  
This design rule can guide the realization of stress focusing and fracturing protection in real materials.

\end{abstract}

\maketitle

\section{\label{sec:intro}Introduction}

How materials respond to applied forces has been a crucial question to understand since ancient times when stable and robust machinery was desired. In the past decades, the role geometry plays in mechanical responses has begun to grasp attention, and a set of materials named mechanical metamaterials, in which the mechanical properties are controlled by geometry, has been developed ~\cite{liu2000locally,Huber2016,Bertoldi2017,Yu2018,Xin2020}. These innovative designs are insensitive to the composition and the scale because of the consistency in geometry, and they offer many interesting engineering opportunities due to their unusual features such as negative Poisson's ratio~\cite{Lakes2017} and negative stiffness~\cite{Nicolaou2012}. Topological mechanical metamaterials (TMMs) represent a special class of metamaterials in which the mechanical responses are governed by the topology of the bulk phonon bands, giving them robust properties immune to imperfections~\cite{Kane2013,Lubensky2015,Mao2018,Sun2020}. TMMs can be realized in many different physical systems~\cite{bilal2017intrinsically,Rocklin2017a,Ma2018,Xiu2022,Jolly2022,Bergne2022,widstrand2023stress,widstrand2023robustness}, opening up opportunities to study fascinating physical phenomena and to engineer intriguing functionalities~\cite{Paulose2015a,Sun2012,Zhang2018,Zhou2020}. A feature of particular interest is the focusing of stress at the domain wall when two topologically distinct domains are connected together~\cite{Paulose2015a,Zhang2018}. The fact that stresses in the structure are focused at the domain wall provides protection against the development of fracturing sites in the structure~\cite{Zhang2018}. 

In the linear approximation, it is understood that a lattice can have both zero energy modes (ZMs), which allow the lattice to deform without energy cost, and states of self-stress (SSSs), which allow the lattice to carry stress without causing net force on the sites~\cite{Kane2013}. ZMs and SSSs are related by the Maxwell-Calladine index, which is calculated from the number of degrees of freedom and the number of constraints in the lattice~\cite{Calladine1978,Sun2012}. 
In topological Maxwell lattices, ZMs and SSSs can be directed to become exponentially localized at lattice edges and interfaces by tuning the unit cell geometry~\cite{Kane2013}.  
The SSSs supply a backbone for the lattices in response to external stresses~\cite{Chen2014} and offers an understanding of the aforementioned stress focusing and fracture protection phenomena~\cite{Zhang2018}. 

Besides the theoretical advancement in TMMs, efforts have been made toward the engineering and manufacturing of TMMs~\cite{bilal2017intrinsically,Rocklin2017a,Ma2018,Bergne2022,Xiu2022,Jolly2022,widstrand2023stress,widstrand2023robustness}. One known difficulty in completing these tasks is the energy cost associated with hinge bending~\cite{bilal2017intrinsically,Guo2019,Bergne2022,Xiu2022}, which is not captured in the theory of TMMs, where free joints are assumed. The non-zero bending stiffness of the hinges has diluting effects on the mechanical properties predicted by theory. It causes an up-shift of the ZM bands from zero frequency,  causing hybridization with bulk bands at long wavelength, which leads to lower contrast between the predicted hard and soft boundaries of the lattice~\cite{Ma2018,stenull2019signatures,Sun2020,saremi2020topological,Jolly2022,charara2022omnimodal}. It  also intensify the stresses at the joints away from where the SSSs are localized, which masks the focusing effect of the stress responses~\cite{widstrand2023stress,widstrand2023robustness}. 
Despite the undesired engineering and manufacturing challenges, the bending stiffness at the joints offers a new perspective for controlling the transformations through which the lattice is able to switch between topological domains~\cite{Rocklin2017a,Xiu2022,Xiu2023}.

In this paper, we report a systematic study of the effect of weak interactions beyond the ideal Maxwell connectivity, such as bending stiffness at the hinges which are ubiquitous in manufactured TMMs, on the topological stress focusing effect in Maxwell lattice TMMs.  Although these weak interactions cause stresses in the bulk at the hinges, diluting the topological stress focusing, we find a general design rule for minimizing stress in the bulk by tuning the unit cell geometry, and we demonstrate this design rule via the agreement between analytic theory and numerical simulations of mechanical loading on these lattices.  Our results will guide practical designs for Maxwell lattice TMMs for stress focusing, fracture control, and impact mitigation.



\section{\label{sec:projSSS} Theory: topologically protected SSS and additional weak interactions}

In this section, we theoretically analyze the effect of weak interactions in non-ideal lattices on topological stress focusing.  
We first review \xm{fundamental theories of topological polarization in ideal Maxwell lattices and how  topological SSSs control stress distribution in Maxwell lattices.  We then show how the addition of weak interactions, in terms of next-nearest neighbor (NNN) bonds, leads to additional SSSs in the bulk that governs the stress distribution the bulk.}


\subsection{Maxwell lattices, topological polarization, and states of self stress}
For a lattice of $N$ sites with equilibrium positions $\mathbf{r}_i$ ($i = 1,\dots,N$) and $N_c$ bonds (here and elsewhere we use bond and spring interchangeably) in $d$ dimensions, the elastic energy can be generally written as
\begin{equation} \label{eq:bondEnergy}
	H = \sum_{\alpha \in \text{bonds}} \frac{k_\alpha}{2} \qty(\mathbf{l}_{i_\alpha j_\alpha} - \mathbf{l}^0_{i_\alpha j_\alpha})^2
\end{equation}
where $i_\alpha$ and $j_\alpha$ are the indices of the sites that bond $\alpha$ connects, $\mathbf{l}^0_{i j} = \mathbf{r}_{i} - \mathbf{r}_{j}$, $\mathbf{l}_{ij} = \mathbf{r}_i - \mathbf{r}_j + \mathbf{u}_i - \mathbf{u}_j$, and $\mathbf{u}_i$ is the displacement of the site $i$. 
Up to quadratic order in displacements $\mathbf{u}_i$, the elastic energy is
\begin{equation} \label{eq: linE}
    H = \frac{1}{2} \mathbf{u}^\intercal \cdot \vb{D} \cdot \mathbf{u}
\end{equation}
where $\mathbf{u}$ is the column vector containing all displacement components of all sites, $^\intercal$ stands for transpose, $\vb{D} = \vb{C}^\intercal \cdot \vb{k} \cdot \vb{C} = \vb{Q} \cdot \vb{k} \cdot \vb{Q}^\intercal$ is the dynamical matrix, in which $\vb{k}$ is the diagonal matrix with springs constants as of all bonds as its entries, $\vb{C}$ is the compatibility matrix that relates the $dN$-dimensional site displacement column vector $\mathbf{u}$ to the $N_c$-dimensional column vector of bond extensions $\mathbf{e}$, and $\vb{Q} = \vb{C}^\intercal$ is the equilibrium matrix that relates the $N_c$-dimensional bond tensions $\mathbf{t}$ to the $dN$-dimensional column vector of forces on sites~\cite{Kane2013}:
\begin{equation} \label{eq:CQmat}
    \vb{C} \cdot \mathbf{u} = \mathbf{e},\ \ \ \ \ \vb{Q} \cdot \mathbf{t} = \mathbf{f}.
\end{equation}
The entries of the $\vb{C}$ matrix are determined by the equation for the extension of the spring $\alpha$
\begin{equation} \label{eq: bextension}
    e_\alpha = \hat{l}_\alpha \cdot \qty(\mathbf{u}_{i_\alpha} - \mathbf{u}_{j_\alpha}),
\end{equation}
where, once again, where $i_\alpha$ and $j_\alpha$ are the indices of the sites that bond $\alpha$ connects, and $\hat{l}_{ij}$ is the unit vector pointing from site $j$ to site $i$. The null space of $\vb{C}$ gives the ZMs as it represents a set of site displacements that cause no bond extensions, and the null space of the $\vb{Q}$ represents SSSs as it represents a set of bond tensions that cause no net force on the sites. 

\begin{figure*}
        \centering
        \includegraphics[width=0.8\textwidth]{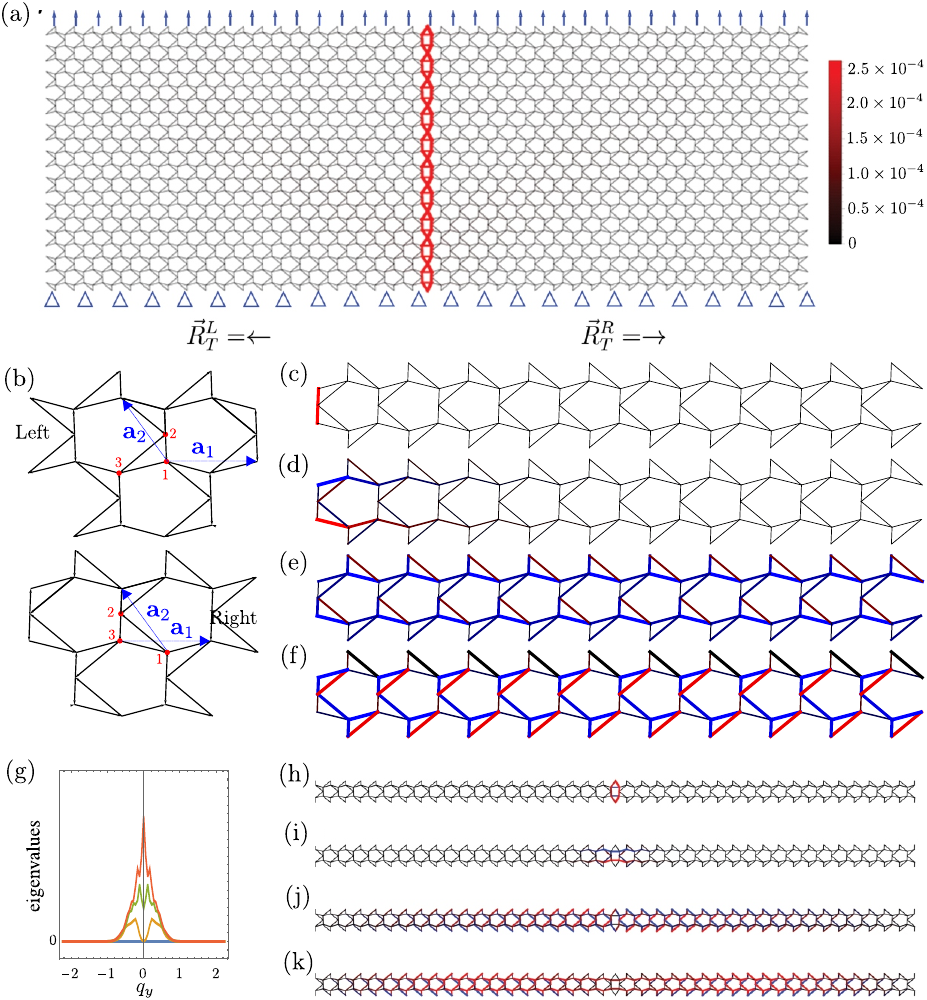}
        \caption{Comparison between simulation and analytic results on an ideal (NN only) Maxwell lattice under a vertical load. (a): An energy minimization simulation by fixing the bottom boundary and giving a vertical displacement on all sites at the top boundary with a magnitude $u = 10^{-3}$. The left and right boundaries are  free, and sliding along $x$ is allowed for the top and bottom boundaries. The triangle marks the fixed boundary and the arrows mark the displaced boundary. The topological polarization vectors $\mathbf{R_T}$ for the left and right domains are shown with arrows. The intensity of the red color and the thickness of the bonds both increase with the magnitude of stress on the bonds. The maximum stress in the response $t_{max} = 2.6\times 10^{-4}$ in units of force. 
        (b): Unit cells of the left and right domains; the red circles are the sites inside the unit cell. The black lines show the bonds in the unit cell. The blue arrows show the lattice vectors.
        (c-f): 4 states of self-stress from unit cell analysis obtained by first solving for the four values of $q_x$ from equation $\det \mathbf{Q}(q_x,q_y = 0) = 0$ corresponding to the unit cell of the right domain, then plugging each of them into the matrix $\mathbf{Q}(q_x,q_y = 0)$ and then obtaining the nullspace in each case. (c), (d), (e), (f) correspond to $q_x = 2\pi \mp 13.8174i, 2\pi \mp 0.739487i,0, 0$, respectively. The SSSs corresponding to the left domain are mirror images of (c-f) w.r.t. a vertical mirror.
        (g): The lowest 4 eigenvalues of $\tilde{\mathbf{D}}(q_y) = \mathbf{C}(q_y)\mathbf{Q}(q_y)$ of the super-cell shown in (h-k). The supercell is Bloch-periodic in $y$-direction and open in $x$-direction. At large values of $|q_y|$ all 4 eigenvalues are zero. At $q_y = 0$ two of them are lifted to nonzero eigenvalue, the other two remain at zero. (h-k): The lowest 4 eigenstates of $\tilde{\mathbf{D}}(q_y=0) = \mathbf{C}(q_y)\mathbf{Q}(q_y)$ from supercell analysis. The red and blue represent tension and compression for the SSS respectively, and the thicknesses are proportional to the magnitudes on the bonds. The modes in (h) and (i) have zero eigenvalue, thus they are SSSs, but the modes in (j) and (k) are elevated to nonzero eigenvalue, hence they are not SSSs. }
        \label{fig:supercell_NN}
\end{figure*}

    
For a homogeneous system with translation symmetry, it is easier to work with the compatibility matrix in Fourier space $\mathbf{C}(\mathbf{q})$, which is $n_c \times nd$ matrix, (where $n_c$ is the number of bonds per unit cell and $n$ is the number of sites per unit cell). As shown in the seminal paper by Kane and Lubensky~\cite{Kane2013}, for a system with a domain wall between two different Maxwell lattices, the number of SSSs localized at a the domain wall can be calculated from the properties of the $\mathbf{C}(\mathbf{q})$ matrices of each Maxwell lattice; below we describe this briefly and discuss the consequences for our specific system shown in Fig.~\ref{fig:supercell_NN}(a).
    
From the compatibility matrix $\mathbf{C}(\mathbf{q})$ of a homogeneous lattice at the Maxwell point (i.e., $n_c = nd$)  in Fourier space, the dynamical matrix can be obtained as $\mathbf{D}(\mathbf{q})=\mathbf{C}^\dagger(\mathbf{q})\mathbf{C}(\mathbf{q})$ (where we chose the spring constants and the masses of the particles to be 1 for simplicity). The normal mode frequencies of the lattice $\omega^2(\mathbf{q})$ are the eigenvalues of the Dynamical matrix $\mathbf{D}(\mathbf{q})$. Kane and Lubensky \cite{Kane2013} defined a `square root' of the dynamical matrix, which in reciprocal space takes the following form:
\begin{equation}
    \mathcal{H}(\mathbf{q}) =\begin{pmatrix}
    \mathbf{0} &  \mathbf{C}^\dagger(\mathbf{q})\\
    \mathbf{C}(\mathbf{q}) &\mathbf{0}
    \end{pmatrix}.
\end{equation}
For every nonzero eigenvalue $\omega^2(\mathbf{q}) $ of $\mathbf{D}(\mathbf{q})$, $\mathcal{H}(\mathbf{q})$ has two eigenvalues $\pm \omega(\mathbf{q})$. The zero modes of $\mathcal{H}(\mathbf{q})$ include nullspace of $\mathbf{C}(\mathbf{q})$ (ZMs) and  nullspace of $\mathbf{C}^\dagger(\mathbf{q})$ (SSSs).
Maxwell-Calladine theorem~\cite{Calladine1978,Lubensky2015} dictates that the number of ZMs ($n_0$) and number of SSSs ($n_s$) are equal ($n_0 = n_s$) for a Maxwell lattice. The matrix $\mathcal{H}(\mathbf{q})$ has the property that $S \mathcal{H}(\mathbf{q}) S = - \mathcal{H}(\mathbf{q})$, where $S =\text{Diag}\{\mathbbm{1},-\mathbbm{1}\}$. This property is known as chiral (or sublattice) (anti)symmetry in the literature. Also, it is easy to check that $\mathcal{H}(\mathbf{q})$ has time reversal symmetry: $\mathcal{H}(\mathbf{q}) = \mathcal{H}^*(-\mathbf{q})$, where $^*$ is complex conjugation. These two symmetries put the matrix $\mathcal{H}(\mathbf{q})$ in BDI class of Altland Zirnbauer (AZ) classification \cite{AltlandZirnbauer}. According to the AZ periodic table~\cite{RevModPhys.88.035005} BDI (or AIII if we exclude time-reversal symmetry) class has a $\mathbbm{Z}$ topological index in 1D. Hence, a topological polarization vector $\mathbf{R}_T$ of a 2D Maxwell lattice can be defined as
\begin{equation}
    \begin{split}
        \mathbf{R}_T &= \sum_{i} n_i \mathbf{a}_i,\\
        n_i &= \frac{1}{2\pi i} \oint_{\mathbf{q}\rightarrow\mathbf{q}+\mathbf{G}_i} d\mathbf{q}\cdot \mathbf{\nabla}_\mathbf{q}\log \det \mathbf{C}^\dagger(\mathbf{q}),
    \end{split}
\end{equation}
where $\mathbf{a}_i$ are the lattice vectors, $\mathbf{G}_i$ are the reciprocal lattice vectors such that $\mathbf{G}_i\cdot \mathbf{a}_j = 2\pi \delta_{ij}$, and $n_i \in \mathbbm{Z}$. Note that this definition of the polarization is only applicable if the $\mathbf{C}(\mathbf{q})$ is a square matrix with $n_c = nd$ or in other words the lattice is Maxwell. Also, we assumed that there are no bulk ZMs (except the uniform translations); hence, the integers $n_i$ do not depend on $\mathbf{q}$. It was shown in \cite{Kane2013} that at a domain wall between two Maxwell lattices the total number of ZMs at the domain wall at every wave number $q$ parallel to the domain wall is given by
\begin{equation}
    \label{Eq:ZMDW}
    \begin{split}
    n_{dw} &= \frac{1}{2\pi}\mathbf{G}\cdot(\mathbf{R}_T^L-\mathbf{R}_T^R),
    \end{split}
\end{equation}
where the superscripts $L$ and $R$ denote systems on the left and right of the domain wall respectively, $\mathbf{G}$ is smallest reciprocal lattice vector pointing from left to right of the domain wall, and we assumed that the polarizations for both systems were calculated in the same gauge. Note that if this number is positive, it counts the number of ZMs, whereas if the number is negative, its absolute value counts the number of SSSs~\cite{Kane2013}. 

We use the example shown in Fig.~\ref{fig:supercell_NN}(a) throughout this section to illustrate our points. Fig.~\ref{fig:supercell_NN}(a) contains a domain wall between two deformed kagome lattices which are mirror images of each other. A unit cell of the lattice on the left (see Fig.~\ref{fig:supercell_NN}(b)) consists of three sites with third and first sites having positions $\mathbf{r}_3^L -\mathbf{r}_2^L = (-0.510684, -0.42569)$ and $\mathbf{r}_1^L -\mathbf{r}_2^L = (0.0105642, -0.297959)$ w.r.t. the second site, whereas in the case of the lattice on the right, the third and first sites have positions $\mathbf{r}_3^R -\mathbf{r}_2^R = (-0.0105642, -0.297959)$ and $\mathbf{r}_1^R -\mathbf{r}_2^R = (0.510684, -0.42569)$ w.r.t. the second site. Both lattices have lattice vectors $\mathbf{a}_1 = (1,0)$ and $\mathbf{a}_2 = (-0.5,0.7)$ (hence reciprocal lattice vectors $\mathbf{G}_1 = 2\pi(1,1/1.4)$, $\mathbf{G}_2 = 2\pi(0,1/0.7)$). The three sites are connected by bonds to each other inside the unit cell. Furthermore, site 1 is connected to site 3 in the next unit cell in $\mathbf{a}_1$ direction, site 2 is connected to site 1 in the next unit cell in $\mathbf{a}_2$ direction, site 3 is connected to site 2 in the next unit cell in $-\mathbf{a}_1-\mathbf{a}_2$ direction (note that there are other ways of choosing bonds, but this choice of bonds makes the unit cell symmetric.  \xm{This unit cell contains 3 sites, leading to 6 degrees of freedom (DOFs).  With the 6 nearest neighbor (NN) bonds, this defines an ideal Maxwell lattice. }

The polarization vectors of the two lattices are $\mathbf{R}_T^L = -\mathbf{a}_1$, $\mathbf{R}_T^R = \mathbf{a}_1$, and the smallest reciprocal lattice vector in the $x$-direction (perpendicular to the domain wall) is $\mathbf{G} = 2\mathbf{G}_1-\mathbf{G}_2$. Consequently $n_{dw} = -4$, which implies that there are 4 SSSs at each wave number $q_y$ parallel to the domain wall ($y$-direction). \xm{The unit cell geometry is chosen arbitrarily.  As in the theory discussed above, as long as the topological polarization is fixed, small changes in the geometry only results in change of the localization length of topological ZMs and SSSs, without altering their site of localization, as guaranteed by the topological protection. }

Since under affine uniaxial strain in $y$-direction the strain energy only depends on the projection of the strain onto the $q_y = 0$ SSSs~\cite{Lubensky2015,Paulose2015a,Zhang2018}, we only focus on those. One can obtain all the information on the SSSs at $q_y = 0$ from the equilibrium matrix $\mathbf{Q}(q_y = 0)$. First, the equation $\det \mathbf{Q}(q_x,q_y=0) = -0.352071 -0.567495 \cos(q_x/2) +0.919566 \cos(q_x) = 0$ has four solutions $q_x = 2\pi \mp 13.8174i, 2\pi \mp 0.739487i, 0, 0$ for the left and the right configurations, respectively. Plugging these solutions into $\mathbf{Q}(q_x,q_y=0)$ and solving for the nullspace of the matrix in each case, we obtain the SSSs shown in Fig.~\ref{fig:supercell_NN}(c-f). From this, the inverse decay lengths of the four SSSs can be read off as $\kappa = 13.8174, 0.739487, 0, 0$, which means that the first two are localized at the domain wall, whereas the last two are extended throughout the system. 
It is also worth noting that the real parts of the first two solutions are $(\Re{q_x},q_y) = (2\pi,0) = \mathbf{G}_1-\frac{1}{2}\mathbf{G}_2 \equiv \frac{1}{2}\mathbf{G}_2$ (modulo a reciprocal lattice vector). The reason for this is that the periodicity along of system along the domain wall is $2\mathbf{a}_2+\mathbf{a}_1 = (0,1.4)$. If we take a unit cell on either side that is compatible with the domain wall, its lattice vectors would be $\mathbf{a}_1' = \mathbf{a}_1$ and $\mathbf{a}_2' = 2\mathbf{a}_2+\mathbf{a}_1$ (this unit cell has twice the area of the primitive unit cell, and has 6 sites inside it), and hence the reciprocal lattice vectors are $\mathbf{G}_1' = \mathbf{G}_1-\frac{1}{2}\mathbf{G}_2$ and $\mathbf{G}_2'=\frac{1}{2}\mathbf{G}_2$, which means the point $\mathbf{q} = \frac{1}{2}\mathbf{G}_2$ in reciprocal space ``folds to'' $\mathbf{q} = \mathbf{0}$ for this choice of the domain wall compatible unit cell.

\subsection{Domain walls, finite system, stress projection, and localization}
In the last subsection we mentioned that among the 4 SSSs at $q_y = 0$, two are extended in $x$-direction. Consequently, the last two are only really SSSs for a system that is infinitely large in $x$-direction. Hence, in a finite system with domain wall such as that in Fig.~\ref{fig:supercell_NN}(a), these two modes are not SSSs. To see this, we do a supercell analysis on the structure shown in {\color{black}Fig.~\ref{fig:supercell_NN}(b-e)} with Bloch-periodic boundary condition in $y$-direction, but open boundary condition in $x$-direction. For this system, we plot the eigespectrum of the matrix $\tilde{\mathbf{D}}(q_y)=\mathbf{C}(q_y)\mathbf{C}^\dagger(q_y)=\mathbf{C}(q_y)\mathbf{Q}(q_y)$ (the supersymmetric partner of the dynamical matrix $\mathbf{D}(q_y)$~\cite{Kane2013}) in {\color{black}Fig.~\ref{fig:supercell_NN}(g)}. The eigenmodes of this matrix with zero eigenvalue are the SSSs. As can be seen from the figure, at large $q_y$, there are 4 zero modes. however at (and near) $q_y = 0$, two of the modes get lifted to nonzero eigenvalues, hence they are not SSSs. In {\color{black}Fig.~\ref{fig:supercell_NN}(b-e)}, the lowest 4 eigenmodes of  $\tilde{\mathbf{D}}(q_y=0)$ are plotted. The ones in {\color{black}Fig.~\ref{fig:supercell_NN}(h-i)} are localized at the domain wall with inverse decay lengths $13.8174 \text{ and } 0.739487$ respectively, and they correspond to zero eigenvalues, hence they remain SSSs. The two in {\color{black}Fig.~\ref{fig:supercell_NN}(j-k)} are extended in $x$-direction, and they have nonzero eigenvalues, these are not SSSs for finite system size with open boundary. 

To understand the role of these two SSSs localized at the domain wall in localizing stress distribution under uniaxial stress in $y$-direction, we adopt the formulation in Refs.~\cite{Lubensky2015,Paulose2015a,Zhang2018}, by projecting the affine strain to the linear space of these SSSs. 
We first note that under strain $\epsilon_{yy}$, $\alpha^{\tiny\text{th}}$ bond gets stretched by amount $e_{\text{aff},\alpha}=\epsilon_{yy}\sin^2\theta_\alpha$, where $\theta_\alpha$ is the angle between $\alpha^{\tiny\text{th}}$ bond and the $x$-direction. Defining $\mathbf{e}_\text{aff}$ as the vector of the affine elongations $e_{\text{aff},\alpha}$, one can show that vector of the bond tensions $\mathbf{t}$ in equilibrium to be
\begin{equation}\label{eq:StrainProjectionF}
    \mathbf{t} = \sum_i \left(\left[\left(\mathbf{K}^{-1}\right)_{ss}\right]^{-1}\mathbf{e}_{\text{aff},s}\right)_i \mathbf{s}_i \xrightarrow{k=1} \mathbf{t} =\sum_i e_{\text{aff},s}^i \mathbf{s}_i,
\end{equation}
where $\mathbf{s}_i$ are the SSSs at $q_y = 0$, $\mathbf{e}_{\text{aff},s}$ is vector of projections $e_{\text{aff},s}^{i}=\mathbf{e}_\text{aff}\cdot\mathbf{s}_i$, $\left(\mathbf{K}^{-1}\right)_{ss} = [\mathbf{s}]^T \mathbf{K}^{-1} [\mathbf{s}]$ where $\mathbf{K}$ is the matrix of bond stiffness and $[\mathbf{s}]$ is the matrix consisting of column vectors $\mathbf{s}_i$. The expression for bond tensions simplifies to $\mathbf{t} =\sum_i e_{\text{aff},s}^i \mathbf{s}_i$ when all bond stiffnesses are equal and taken to be 1. Now, the localized SSS in Fig.~\ref{fig:supercell_NN}(i) has both tensions and compressions (red and blue colored bonds, respectively); hence the projection $e_{\text{aff},s}^i$ of an affine uniaxial strain on it is negligible compared to the projection of the affine uniaxial strain in the SSS in Fig.~\ref{fig:supercell_NN}(h). This is why 
bond tension distribution (or stress distribution) of the finite system under affine strain $\epsilon_{yy}$ should closely resemble the SSS in Fig.~\ref{fig:supercell_NN}(h); since the SSS in Fig.~\ref{fig:supercell_NN}(b) is highly localized at the domain wall (recall that its inverse decay length is $13.8174$), the stress distribution should be highly localized at the domain wall. 

We perform a numerical simulation on a finite system shown in Fig.~\ref{fig:supercell_NN}(a) (see Sec.~\ref{sec:model} figure caption for details) to compare with the super-cell analysis. The stress distribution in the simulation agrees well with the SSS shown in Fig.~\ref{fig:supercell_NN}(h), as we expect from the projection discussed above.

\subsection{Addition of weak next nearest neighbor bonds}  

\begin{figure*}
    \centering
    \includegraphics[width = 0.8\textwidth]{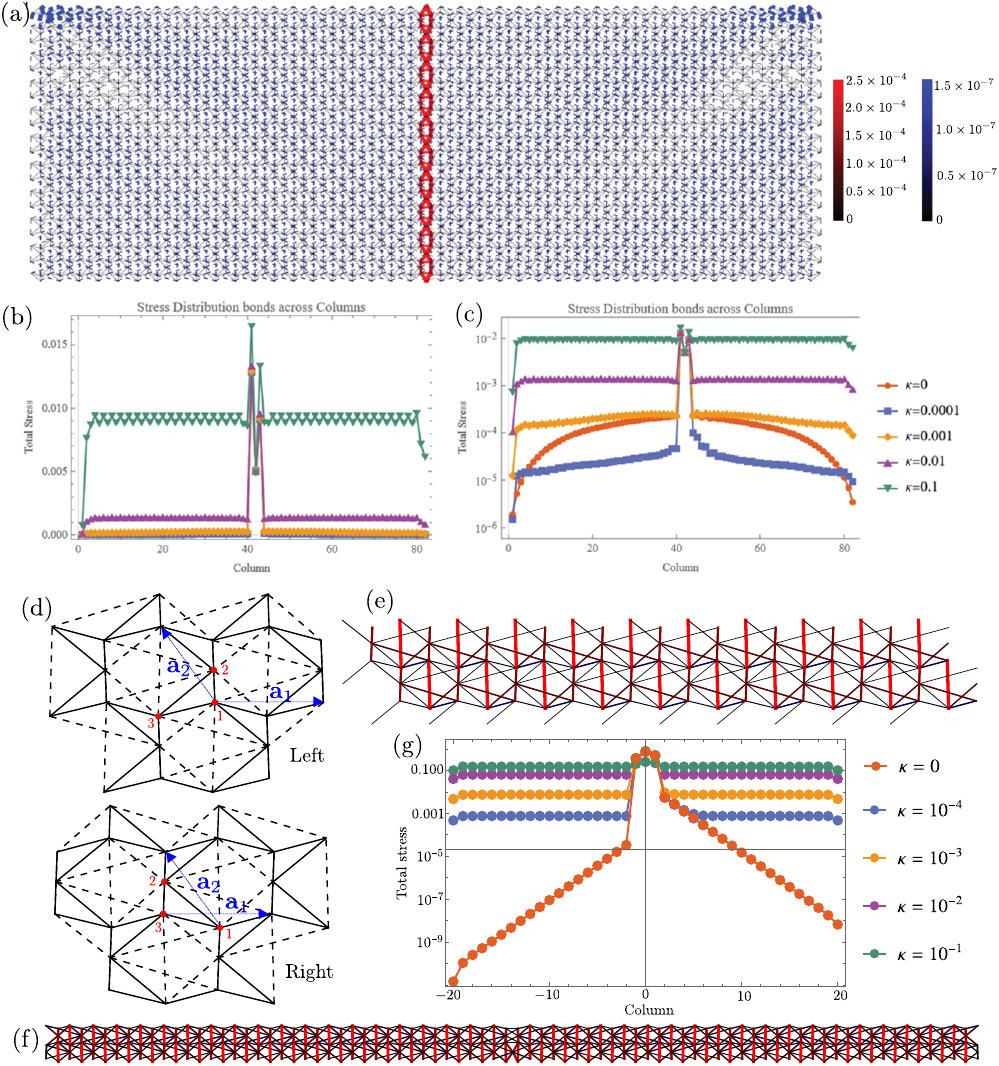}
    \caption{Comparison between simulation and the analytical results on a non-ideal (with NNN bonds) Maxwell lattice under a vertical load. (a): An energy minimization simulation as the one described in Fig. \ref{fig:supercell_NN}(a), but with the inclusion of NNN spring constant $\kappa = 10^{-4}$. The maximum tension on the NN bonds is $t_{max}^{NN} = 2.55\times 10^{-4}$ in the units of force, and $t_{max}^{NNN} = 1.56\times10^{-7}$ on the NNN bonds. The NN bonds are labeled in red, while the NNN bonds are represented in blue. The intensity of the color increases with the stress magnitude. (b,c): The total stress on bonds across columns in the lattice in linear (b) and log (c) scales, at different  $\kappa$. The peak in the middle correspond to the SSS domain wall. 
    (d): Unit cells of the left and right domains; the dark red circles are the sites inside the unit cell, the light red circles are the sites at the adjacent unit cell. The black solid and dashed lines show the NN and NNN bonds in the unit cell, respectively.
    (e): Projection of affine strain $\epsilon_{yy}$ onto the $6$ new SSSs at $\mathbf{q} = \mathbf{0}$ due to the additional NNN bonds for a bulk system with unit cell the same as on the right of the domain wall. Projection of affine strain $\epsilon_{yy}$ onto the $6$ new SSSs for the bulk system on te left of the domain wall is just a mirror image of this under a vertical mirror. Here, we take $\kappa/k = 1$.
    (f): Bond tensions in the supercell under affine strain $\epsilon_{yy}$ obtained using Eq.~\eqref{eq:StrainProjectionF} by projecting the strain onto the $474$ SSSs at $q_y = 0$ for the supercell (we take Bloch-periodicity in the $y$-direction and open boundary condition in the $x$-direction). Away from the the domain wall the bond tensions closely resemble that in (e). Here, we take $\kappa/k = 1$ to exaggerate the bond tensions in the NNN bonds.
    (g): Stress in unit cells plotted across the columns from the supercell analysis at different $\kappa$, which is analogous to (c) from the numerical simulation.}
        \label{fig:NNN_sim}
\end{figure*}

Real specimen of manufactured TMMs are never exactly at the Maxwell point, which would require perfectly flexible hinges.  To quantify the effect of hinge bending stiffness on the stress localization in topologically polarized Maxwell TMMs, we add weak NNN bonds to the Maxwell lattices, which to leading order represents the angular bending stiffness at the hinges, and study the stress distribution analytically and computationally. 
In particular, the extensions/compressions of these NNN bonds are equivalent to bending at the hinges/sites in real metmaterials, and  stress concentration due to bending at the hinges in the bulk of the lattice has been observed computationally~\cite{widstrand2023stress}. 

With the addition of the $6$ NNN bonds per unit cell (Fig.~\ref{fig:NNN_sim}(d)), the number of constraints in each unit cell is 6 more than the number of degrees of freedom. As a consequence, for the homogeneous system on either side of the domains the matrix $\mathbf{Q}(\mathbf{q}) \equiv \mathbf{C}^\dagger(\mathbf{q})$ is $6\times 12$, which implies that there are $12-6=6$ extra SSSs at every $\mathbf{q}$ than before adding NNN bonds. \xm{Although these SSSs arise due to NNN bonds, the tensions associated with these SSSs lie on both NN and NNN bonds.}

Similar to the discussion at Eq.~\eqref{eq:StrainProjectionF}, the projection of the affine strain to these $\mathbf{q} = \mathbf{0}$ SSSs determines the stress distribution in the bulk of the lattice far from the domain wall.  
At $\mathbf{q} = \mathbf{0}$, there are two more SSSs (in total $6+2 = 8$) due to the fact that there are two ZMs that are uniform translations. However, there were already 2 SSSs for the system with only NN bonds (the two extended ones shown in Fig.~\ref{fig:supercell_NN}), which are still SSSs of the system with both NN and NNN bonds. These two SSSs do not remain SSSs for a finite system as discussed earlier. Hence, we are only concerned with the 6 new SSSs at $\mathbf{q} = \mathbf{0}$ due to the NNN bonds and the projection of the affine strain onto them for the calculation of stress in the bulk. If the projections of the affine strain onto these SSSs are nonzero, there will be finite extension/compression of bonds away from the domain wall as we show in Fig.~\ref{fig:NNN_sim}(e). 
\xm{It is worth noting that the $\mathbf{K}$ matrix is no longer an identity matrix here: the NN bonds stiffness $k$ and NNN stiffness $k_{NNN}$ are typically different. }
Furthermore, since these new SSSs are due to the additional NNN bonds, the NNN bonds would get extended/compressed under strain, which implies stress concentration at the hinges in the bulk away from the domain wall in real metamaterials.

To show the fact that a linear combination of the 6 bulk SSSs at $\mathbf{q} = \mathbf{0}$ are responsible for the NNN bond stretching/compression under affine uniaxial strain $\epsilon_{yy}$, we first perform a supercell analysis. To this end, we create a supercell with a domain wall between two homogeneous systems (as shown in Fig.~\ref{fig:NNN_sim}(f)) with $N_0 = 20$ domain wall compatible unit cells (note that a domain wall compatible unit cell has twice as many DOFs as a primitive unit cell) on each side of the domain wall. In total there are $6\times2N_0+1 = 241$ sites (482 DOFs) and $954$ bonds. We apply the open boundary condition in the $x$-direction and Bloch-periodic boundary condition in the $y$-direction. Due to the excess of bonds ($954-482=472$), at every wave number $q_y$ parallel to the domain wall there are 472 bulk SSSs. At $q= 0$, the number of SSSs is $474$ (the extra two are due to translation symmetry). Among these $474$ SSSs, $2$ are localized at the domain wall (these were already present in the nearest neighbor model without the NNN bonds), and $472$ are bulk SSSs appearing due to excess bonds at every unit cell. In principle, the affine uniaxial strain can have projections onto all of these $474$ SSSs. However, if our argument in the last paragraph is correct, among these $472$ extra SSSs, only $6$ of them carry stress deep inside the bulk away from the domain wall and the edges. To confirm this, we plot the tension obtained using Eq.~\eqref{eq:StrainProjectionF} the projection of affine strain onto the 474 SSSs from the supercell analysis in Fig.~\ref{fig:NNN_sim}(f), the same obtained from the projection of affine strain onto the 6 bulk SSSs obtained from unit cell analysis at $\mathbf{q}=0$ discussed in the last paragraph in Fig.~\ref{fig:NNN_sim}(e) (in both cases, we take the bond stiffnesses for both NN and NNN bonds to be 1); the bond tension patterns in these two figures match quite well deep inside the bulk.

Furthermore, we perform a numerical simulation on a finite system with NNN bonds, as shown in Fig.~\ref{fig:NNN_sim}(a), where an axial displacement is applied (see Sec.~\ref{sec:model} for details).  
As can be seen from the color scale in Fig.~\ref{fig:NNN_sim}(a), there is stress localization near domain wall on the NN bonds (which is very close to what we found in Fig.~\ref{fig:supercell_NN}(a) for the NN only lattice), but NNN bonds are stretched/compressed more or less homogeneously in the bulk away from the domain wall, which is what we predicted above using the SSS argument. We also plot the total tension on bonds in each column in the lattice in Fig.~\ref{fig:NNN_sim}(b-c) from the full simulation for different values of the ratio $\kappa/k$ of the NNN bond stiffness and NN bond stiffness. Notice that for each value of $\kappa/k$, the total bond tension is almost independent of the column number in the bulk away from the domain wall and the edges. Moreover,  this total bond tension increases linearly with increasing $\kappa$ for small $\kappa/k$. This can be understood using Eq.~\eqref{eq:StrainProjectionF} in the following way. The bond tension in equilibrium is given by $\mathbf{t} = \sum_i \left(\left[\left(\mathbf{K}^{-1}\right)_{ss}\right]^{-1}\mathbf{e}_{\text{aff},s}\right)_i \mathbf{s}_i$. The homogeneous bond tensions in the bulk away from the domain wall is due to SSSs that appear due to the extra NNN bonds. For these SSSs, $\left[\left(\mathbf{K}^{-1}\right)_{ss}\right]^{-1} \propto \kappa$ for small $\kappa/k$; hence $|\mathbf{t}| \propto \kappa$. To confirm this, we plot $\mathbf{t} = \sum_i \left(\left[\left(\mathbf{K}^{-1}\right)_{ss}\right]^{-1}\mathbf{e}_{\text{aff},s}\right)_i \mathbf{s}_i$ for different values of $\kappa/k$ under affine strain $\epsilon_{yy}$ in the supercell analysis in Fig.~\ref{fig:NNN_sim}(g). We find the same behavior of the total bond tensions in each column as a function of $\kappa$ as we found in full numerical simulation in Fig.~\ref{fig:NNN_sim}(c).

\section{\label{sec:model}Simulation models}
In this section we discuss two simulation models we use to study stress distribution in topological Maxwell TMMs.  The first model is the NNN bond model which we introduced in the analytic theory, where NNN bonds are added to the lattice capturing the bending stiffness at the hinges.

The second model is the angular spring model (AS). The model associates an energy cost to the angular change from the rest angles with quadratic form
\begin{equation} \label{eq: ASEnergy}
    H_{\theta} = \sum_{i,jk} \frac{\kappa_{\theta_{i,jk}}}{2}\qty(\theta_{i,jk} - \theta^0_{i,jk})^2
\end{equation}
where $\kappa_{\theta_{i,jk}}$ and $\theta^0_{i,jk}$ are the angular spring constant and the rest angle for angle $\theta_{i,jk}$ formed by the central site $i$ and its neighboring sites $j,k$. From the AS energy function, the forces on the particles $i,j,k$ from each angle 
        \begin{equation} \label{eq: ASForce}
            F_{l\alpha}=-\sum_{\{i,jk\}} \kappa_{\theta_{i,jk}} \qty( \theta_{i,jk}-\theta_{i,jk}^0) \frac{\partial \theta_{i,jk}}{\partial x_{l\alpha}},
        \end{equation}
\xm{where $\alpha$ denotes $x,y$ directions,}
can be calculated by taking the first derivative of Eq.~\eqref{eq: ASEnergy} to reflect the bending stresses in the lattice. A detailed full expression of the AS forces and the angle convention we use can be found in the App.~\ref{app:AS}. In our simulations, \xm{angular springs are added at all sites for bond pairs that do not belong to the same triangle (the bending of in-triangle bond pairs already cost energy in the NN-only lattice),} and
$\kappa_{\theta_{i,jk}}$ is taken to be a constant for all angular springs. 

The NNN and the AS model capture very similar effects where additional constraints are added to the ideal Maxwell lattice to constrain DOFs that were floppy.  In deed, to leading order, the AS stiffness
is related to the NNN spring stiffness $\kappa$ by a geometric parameter which is determined by the geometry of the unit cell, as it shows how much an NNN spring changes in length as the corresponding angle changes:
        \begin{equation} \label{eq:k_conversion}
            \kappa = -\frac{l_{ij} l_{ik} \sin \theta}{\sqrt{l^2_{ij} + l^2_{ik} -2 l_{ij} l_{ik} \cos \theta}} \kappa_{\theta}.
        \end{equation}

\begin{figure*}
        \centering
        \includegraphics[width=0.95\textwidth]{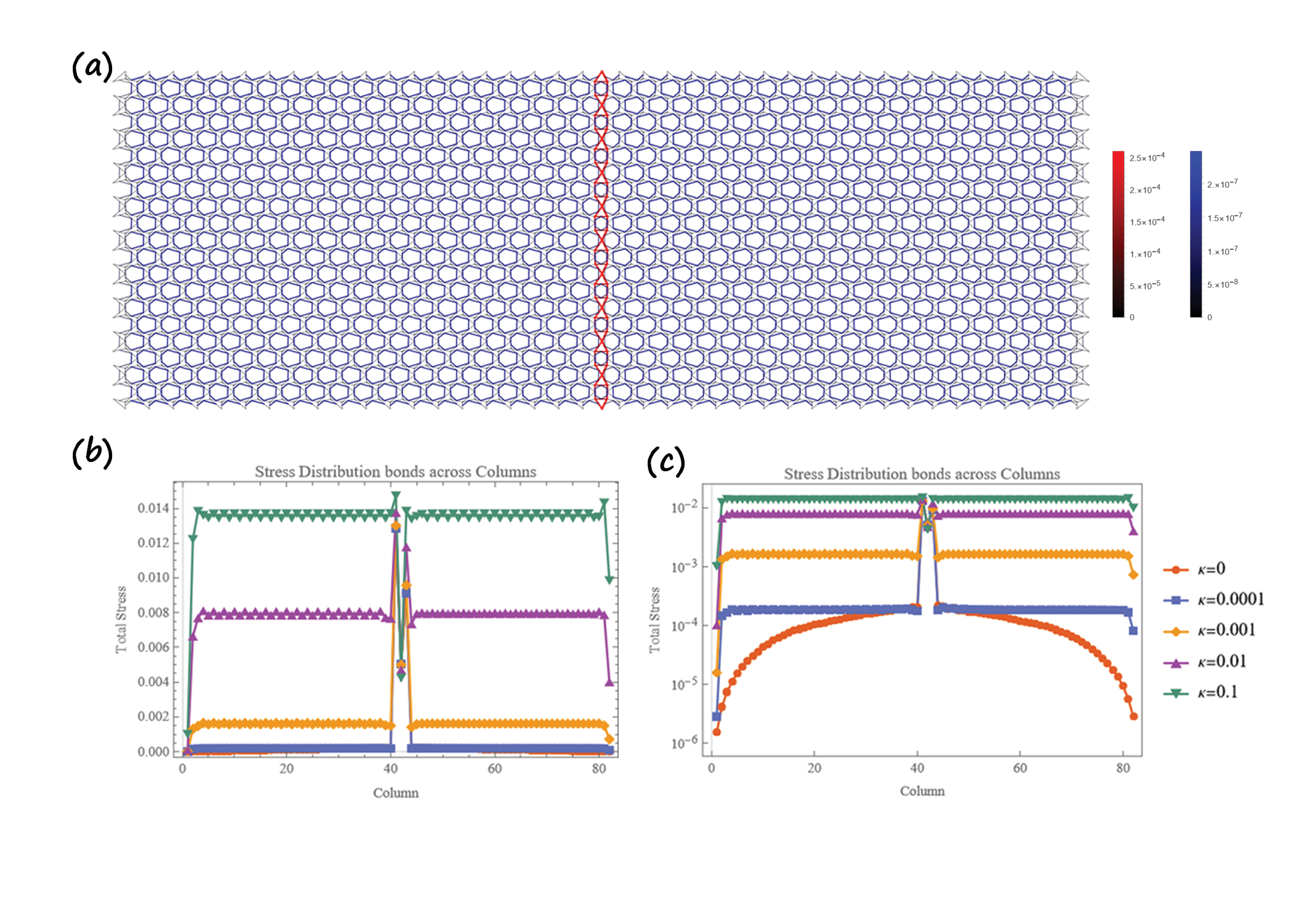}
        \caption{AS model under a vertical load (a): An energy minimization simulation as the one described in Fig. \ref{fig:supercell_NN}(a), but with AS with spring constant $\kappa_{\theta} = 10^{-4}$ instead of NNN bonds. The maximum tension on the NN bonds is $t_{max}^{NN} = 2.59\times 10^{-4}$ in the units of force, and $t_{max}^{AS} = 2.48\times10^{-7}$ on the AS. The NN bonds are labeled in red, while the AS are represented in blue (as a short line connecting the centers of the bond pairs). The intensity of the color increases with the stress magnitude. (b,c): The total stress on bonds across columns in the lattice in linear (b) and log (c) scales at different AS stiffness $\kappa$, where the peak in the middle correspond to the stress focusing at the domain wall. 
        }
        \label{fig:AS_sim}
    \end{figure*}

In the simulation, an axial displacement of $10^{-3}\ll 1$ in $y$-direction is applied to all sites at the top boundary (which corresponds to strain $\epsilon_{yy} \approx 7\times 10^{-5}$), and sites on the bottom boundary are pinned in the $\hat{y}$-direction, while the left and right boundaries are left open, and the sites on the top and bottom boundaries are allowed to slide in $x$. We minimize the energy of the lattice under this boundary condition (for details see App.~\ref{app:NMethod}).  Results from this protocol applied on the NNN model and the AS model are shown in Fig.~\ref{fig:NNN_sim} and \ref{fig:AS_sim}, respectively.

\xm{In both cases, we observe similar effects where at small $\kappa$, sharp stress focusing occurs at the SSS domain wall.  As $\kappa$ increases, this topological focusing is diluted, with stress in the bulk growing linearly with $\kappa$.  This feature, as well as the detailed profile of stress in the unit cell in the bulk, are well captured by our analytic theory based on unit cell analysis and supercell analysis.}


    
Although the AS model is closer to the actual bending stiffness in real TMMs, its multi-body nature makes it more complicated to study analytically.  Thus, we chose to use the NNN model, which produces very similar results, in our analytic theory study in Sec.~\ref{sec:projSSS}.

\section{\label{sec:lattice_design} Design rule for minimizing stress in the bulk}
\begin{figure*}
    \centering
    \includegraphics[width = 0.8\textwidth]{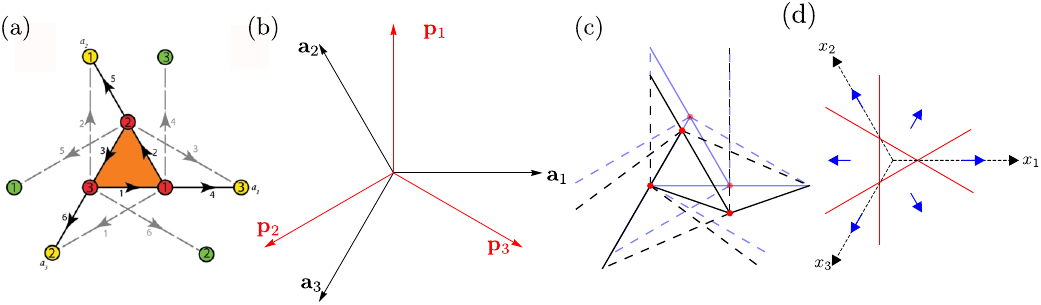}
    \caption{Generalized kagome lattice. (a): Unit cell of the kagome lattice, the black solid lines represent the NN bonds and the gray dashed one the NNN bonds. The red sites are the sites in this unit cell and the green and yellow ones represent the same sites in the neighboring unit cells.
    (b): Lattice vectors $\mathbf{a}_1$, $\mathbf{a}_2$ and $\mathbf{a}_3 = -\mathbf{a}_1-\mathbf{a}_2$. The vectors $\mathbf{p}_1$, $\mathbf{p}_2$ and $\mathbf{p}_3$ are unit vectors rotated counterclockwise by $\pi/2$ from $\mathbf{a}_1$, $\mathbf{a}_2$ and $\mathbf{a}_3$ , respectively.
    (c): The deformation of the kagome lattice when $x_1>0, x_2=x_3=0$. Site 1 and site 2 move by amount $-\sqrt{3}\mathbf{p}_1$ and $x_1\mathbf{a}_3$ under this deformation. The blue and black lines are the bonds before and after the deformation respectively, whereas the solid and dashed lines show NN and NNN bonds respectively. 
    (d): Ternary phase diagram showing topological polarization $\mathbf{R}_T$ for a particular value of $x_1+x_2+x_3>0$. The red lines correspond to one of the $x_i$ being zero. The configurations inside the red triangle are unpolarized. The polarization $\mathbf{R}_T$ in the six sectors surrounding the red triangle are shown by blue arrows.
    }
    \label{fig:GKL_UC}
\end{figure*}
In Sec.~\ref{sec:projSSS}, we showed that addition of NNN bonds (equivalent to having bending energy cost at the hinges) causes stress concentration on NNN bonds away from the domain wall inside the bulk. Now, a natural question arises: how do we minimize the stress in the NNN bonds in the bulk away from the domain wall, \xm{and maintain strong stress focusing at SSS domain walls in these non-ideal Maxwell lattices? This question is crucial as stress in the bulk at carried by bending load on the thin hinges, making the lattice prone to damage.  }
The key observation in this direction was made in Sec.~\ref{sec:projSSS}C, that \emph{the stress concentration on the NNN bonds in the bulk is due to the projection of the affine strain $\epsilon_{yy}$ onto the additional 6 bulk SSSs at $\mathbf{q} = \mathbf{0}$.} This gives the design rule: 

\paragraph*{\underline{Design-rule:}}\textit{To minimize the stress concentration on the NNN bonds, we need to design a deformed kagome lattice such that projection of $\epsilon_{yy}$ on the $6$ extra SSSs at $\mathbf{q} =\mathbf{0}$ due to the additional NNN bonds is minimized keeping the topological polarization intact}. 

In this section, we first review the parametrization of generalized kagome lattice introduced in Ref.~\cite{Kane2013}, and then search for the bulk-stress minimizing configuration in this parameter space using the design rule mentioned above.

\subsection{\label{subsec:GKL} Generalized Kagome Lattice}
In the regular (undeformed) kagome lattice the positions of the three sites in the unit cell are $\mathbf{r}_1 = \mathbf{a}_1/2$, $\mathbf{r}_2 = -\mathbf{a}_3/2$, $\mathbf{r}_3 = \mathbf{0}_3$ where $\mathbf{a}_n = \left(\cos(2\pi(n-1)/3),\sin(2\pi(n-1)/3)\right)$ are the lattice vectors (see Fig.~\ref{fig:GKL_UC}(b)). The unit cell, the sites and the NN and NNN bonds for regular kagome lattice are shown in Fig.~\ref{fig:GKL_UC}(a). Now, this lattice can be deformed in infinitely many ways. However, while deforming the kagome lattice, if we keep the lattice vectors and one of the sites fixed, there are only four degrees of freedom to deform the lattice which are the $x$ and $y$ coordinates of two of the sites, since we can always fix one of the sites. Kane and Lubensky~\cite{Kane2013} used a different parametrization of these deformations (of course, the number of of degrees of freedom of deformation is still 4), in which the expression of the topological polarization as a function of the parameters become simple. In this parametrization, the sites of the deformed kagome lattice are given by
\begin{equation}
    \mathbf{R}_i = \mathbf{r}_i - \sqrt{3}x_i \mathbf{p}_i +x_{i-1} \mathbf{a}_{i+1} + \frac{z}{\sqrt{3}}\mathbf{p}_{i-1},
\end{equation}
where $\mathbf{p}_n = (-\sin(2\pi(n-1)/3), \cos(2\pi(n-1)/3))$ are unit vectors rotated counterclockwise from $\mathbf{a}_n$ by angle $\pi/2$. The effect of $x_1>0, x_2 = x_3 = 0$ is shown in Fig.~\ref{fig:GKL_UC}(c), the effect of the $x_2$ and $x_3$ are similar. Parameter $z$ uniformly dilates the triangle formed by the three sites. In this parametrization the, the expression for the topological polarization is
\begin{equation}
    \mathbf{R}_T=\frac{1}{2}\sum^{3}_{i=1} \text{sgn}(x_i) \mathbf{a}_i,
\end{equation}
which is shown in the ternary phase diagram in Fig.~\ref{fig:GKL_UC}(d). Since we want the domain wall to be vertical (along $y$-direction) and the number of SSSs at every $q_y$ to be $|n_{dw}| = 4$, we want $\mathbf{R}_T$ on the left domain to be $\mathbf{R}_T  =  -\mathbf{a}_1$ which implies $(x_1<0, x_2>0, x_3>0)$, which is the middle left region in the ternary phase diagram.

\subsection{\label{subsec:stress_design}Minimizing bulk stress}
\begin{figure*}
    \centering
    \includegraphics[width = 0.8\textwidth]{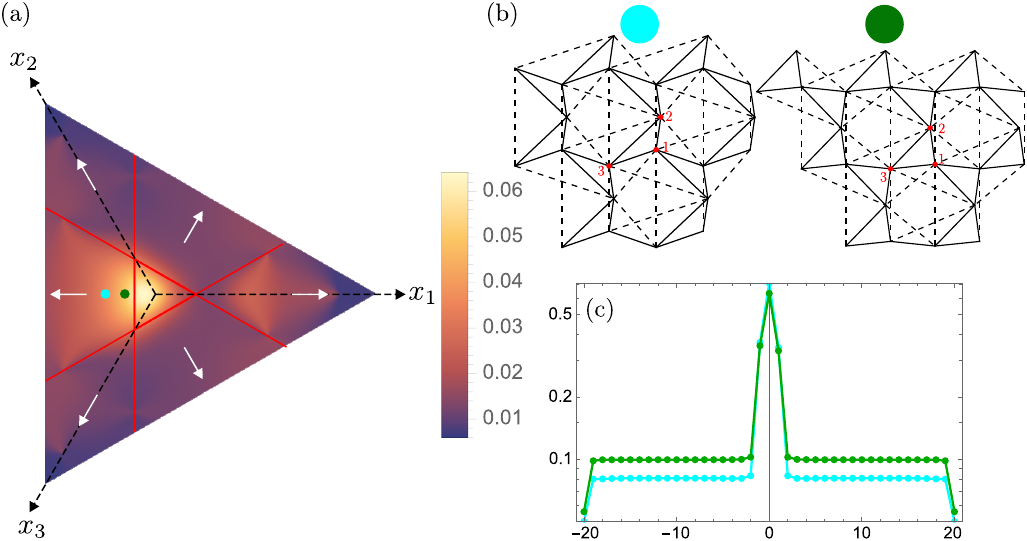}
    \caption{Minimization of stress on the NNN bonds in the bulk. (a): Colormap of the 2-norm of NNN bond tensions in the $(x_1,x_2,x_3)$ ternary phase diagram for $x_1+x_2+x_3 = 0.15$ and  $z= 0$ obtained using the equation $\mathbf{t} = \sum_{i=1}^6 \left(\left[\left(\mathbf{K}^{-1}\right)_{ss}\right]^{-1}\mathbf{e}_{\text{aff},s}\right)_i \mathbf{s}_i$ where $\mathbf{s}_i$ are the 6 extra SSSs at $\mathbf{q} = \mathbf{0}$ due to the additional NNN bonds. Here we used $\kappa/k = 0.01$ and $\epsilon_{yy}=1$. At each red line the coordinate perpendicular to the red line is 0 (for example, $x_1 = 0$ along the vertical red line). The white arrows show the topological polarization in different regions of the parameter space.
    (b): The unit cells corresponding to the cyan and green dots in panel (a). The values of the parameters are $x_1 = -0.1, x_2 = 0.125, x_3 = 0.125, z = 0$ for cyan and $x_1 = -0.03, x_2 = 0.09, x_3 = 0.09, z = 0$ for green.
    (c): Stress in unit cells  (for $\kappa/k = 0.01$) plotted across the columns from the supercell analysis, for the configurations shown in panel (b).
    }
        \label{fig:StressMinimization}
\end{figure*}
Now that we have the design rule and a parametrization of the deformed kagome lattice where the expression topological polarization is simple, we can do a parameter sweep, and for each value of the set of parameters calculate the NNN bond tensions using the equation $\mathbf{t} = \sum_{i=1}^6 \left(\left[\left(\mathbf{K}^{-1}\right)_{ss}\right]^{-1}\mathbf{e}_{\text{aff},s}\right)_i \mathbf{s}_i$ where $\mathbf{s}_i$ are the 6 extra SSSs at $\mathbf{q} = \mathbf{0}$ of the unit cell due to the additional NNN bonds. Since the parameter space is 4-dimensional, visualization of the results is not simple. Hence, we restrict ourselves to $x_1+x_2+x_3 =0.15$ and $z=0$, these constraints make the parameter space 2-dimensional. \xm{This is an arbitrary choice just to demonstrate the optimization.  More thorough parameter sweep can be done following the same equations for practical designs. }

We show the result in Fig.~\ref{fig:StressMinimization}(a). For a left domain with $\mathbf{R}_T = -\mathbf{a}_1$, we look at the central left region (the unit cells of the right domain are mirror images of the unit cell of the left domain). As can be seen from the colormap in Fig.~\ref{fig:StressMinimization}(a), the stress in the NNN bonds decreases as one goes toward left from the vertical red line (along which $x_1 = 0$). We show the unit cells in Fig.~\ref{fig:StressMinimization}(b) at two different points in the parameter space marked by cyan and green dots in Fig.~\ref{fig:StressMinimization}(a). We found 2-norm $|\mathbf{t}|$ of the NNN bond tensions ($|\mathbf{t}|= (\sum_{i=1}^6 t_i^2)^{1/2}$, where the sum is over the 6 NNN bonds) are $0.0357292$ and $0.0535986$ at the cyan and green points (note that $\kappa/k = 0.01$ and $\epsilon_{yy}=1$). Furthermore, the maximum tensions in the NNN bonds are $0.0201862$ and $0.030927$ at the cyan and green points. 

To verify these predictions, next we create supercells for these two configurations analogous to those in Figs.~\ref{fig:supercell_NN} and~\ref{fig:NNN_sim} and evaluate the bond tensions using the formula $\mathbf{t} = \sum_{i=1}^{474} \left(\left[\left(\mathbf{K}^{-1}\right)_{ss}\right]^{-1}\mathbf{e}_{\text{aff},s}\right)_i \mathbf{s}_i$, where $\mathbf{s}_i$ are all the 474 SSSs at $q_y = 0$ of the supercell and plot their norm per unit cell in Fig.~\ref{fig:StressMinimization}(c). Clearly, there is stress concentration at the domain wall, which is expected due to the two domain wall localized topological SSSs. Also, due to the extra SSSs from the NNN bonds, the bond tensions become constant away from the domain wall. However, the bond tensions for the cyan configuration is lower than the green configuration. This matches with our prediction from unit cell strain projection analysis in Fig.~\ref{fig:StressMinimization}(a). Also, the maximum NNN bond tensions for cyan and green configurations are $0.0214706$ and $0.0316642$ respectively, which are in very good agreement with our unit cell results reported earlier, \xm{verifying that the unit cell SSS projection in non-ideal Maxwell TMMs serves as a efficient and accurate prediction of stress in the bulk.}

This brings us to the main findings of the article. We first argued that the hinges in real metamaterials can be modeled as NNN bonds. Then using this model we showed that the fact that there is stress concentration at the hinges away from the domain wall inside the bulk in real metamaterials can be attributed to the extra bulk SSSs coming from the extra NNN bonds. Using this insight, we gave a design rule for Maxwell lattice to minimize the hinge stresses away from the domain wall.

In principle, one can go even further to the left in the ternary phase diagram in Fig.~\ref{fig:StressMinimization}(a) for even smaller stress projection onto the NNN bonds; however, at some point the system becomes self-intersecting (the NN bonds are physical, they can not intersect each other), which is unphysical.

\section{\label{sec:disc}Discussion}
\xm{In this paper, we study the effect of weak interactions beyond the Maxwell constraint-counting on the topological stress focusing effect in Maxwell TMMs, and provide a design rule to minimize the diluting effect of these weak interactions on  topological stress focusing.  }

\xm{
Ideal Maxwell lattices are known to focus stress when the system is under external load, where stress exponentially localizes at SSS domain walls, leaving the bulk virtually stress-free and protected against damage.  
However, in real manufactured TMMs, weak interactions are always present, and are known to lead to stress in the bulk, causing bending loads at hinges, making them prone to damage.}

\xm{
We show that such stress in the bulk is well captured by a simple unit-cell calculation based on the  projection of the load to the SSSs caused by these additional constraints in the bulk, and propose a design rule to optimize the stress focusing of these TMMs, minimizing such stress in the bulk by tuning the unit cell geometry.  We test our theory using simulations of real-space non-ideal Maxwell TMMs under load using two types of models based on NNN bonds and AS springs, and show the agreement between the analytic theory and the simulation.
}

\xm{This design rule provides an efficient and accurate way to find unit cell geometries that optimize TMMs for stress focusing. Such stress focusing effect can potentially be applied to many technologies from fracture protection, stimuli response, and impact mitigation.}

\begin{acknowledgments}
This work is supported by the National Science Foundation NSF-CMMI-2026794 (H.L., S.G., and X.M.)
 and Office of Naval Research MURI N00014-20-1-2479 (H.L., S.S., N.A. E.S. and X.M.).

\end{acknowledgments}

\appendix
\begin{widetext}

\section{Angular Spring Model Energy and Forces}\label{app:AS}
Energy from the angular spring differs from usual springlike interactions because it involves three sites, instead of being a pairwise interaction. We represent these triples with $\{i,jk\}$ where $i$ represents the central particle and $j$ and $k$ represent the side particles. A diagram of the angle composed by central particle $0$ and particles $x$ and $y$ is shown in Fig.~\ref{fig:k_conversion}.


The energy of bending can be written as a springlike interaction on this angle:
\begin{equation}
U=\sum_{\{i,jk\}} \frac{\kappa_{i,jk}}{2}\left( \theta_{i,jk}-\theta_{i,jk}^0\right)^2
\end{equation}
where $\theta_{i,jk}$ and $\theta_{i,jk}^0$ are the current and rest angles associated with the triple $\{i,jk\}$ and $\kappa_{i,jk}$ is the angular stiffness of this bending bond.

This angle can be measured from the positions of the particles with $\theta_{i,jk}=\arccos(\hat{r}_{ij}\cdot \hat{r}_{ik})$ where $\hat{r}_{ij}$ and $\hat{r}_{ik}$ are the unit vectors of the two bond vectors that form the angle $\theta_{i,jk}$. As discussed, this suffers from a limited range of $0$ to $\pi$. We can investigate reflex angles via utilization of the cross product as well as the dot product:
\begin{equation}
\theta_{i,jk} = \pi+\text{sgn}\left(\overrightarrow{r}_{ij} \times \overrightarrow{r}_{ik} \right) \left(\arccos(\hat{r}_{ij}\cdot \hat{r}_{ik})-\pi \right).
\end{equation}
This measure has a range from $0$ to $2 \pi$. Therefore, $\left( \theta_{i,jk}-\theta_{i,jk}^0\right)$ ranges from $-\theta_{i,jk}^0$ to $\left( 2 \pi-\theta_{i,jk}^0\right)$. It is also important to note that $\theta_{i,jk}=2\pi-\theta_{i,kj}$.

The force on each particle from the AS via the first derivative of the energy is:
\begin{equation}
F_{l\alpha}=-\sum_{\{i,jk\}} \kappa_{i,jk} \left( \theta_{i,jk}-\theta_{i,jk}^0\right) \frac{\partial \theta_{i,jk}}{\partial u_{l\alpha}}.
\end{equation}
This derivative of the angle with respect to particle motion is given by
\begin{equation}
\frac{\partial \theta_{i,jk}}{\partial u_{l\alpha}}=-\frac{\textrm{sgn}\left(\overrightarrow{r}_{ij} \times \overrightarrow{r}_{ik} \right)}{\sqrt{1-(\hat{r}_{ij}\cdot \hat{r}_{ik})^2}} \left( \frac{\delta_{jl}-\delta_{il}}{r_{ij}} (\hat{r}_{ik,\alpha} -(\hat{r}_{ij}\cdot \hat{r}_{ik})\hat{r}_{ij,\alpha} ) + \frac{\delta_{kl}-\delta_{il}}{r_{ik}} (\hat{r}_{ij,\alpha} -(\hat{r}_{ij}\cdot \hat{r}_{ik})\hat{r}_{ik,\alpha} ) \right).
\end{equation}

\begin{figure}
    \centering
    \includegraphics[scale = 0.3]{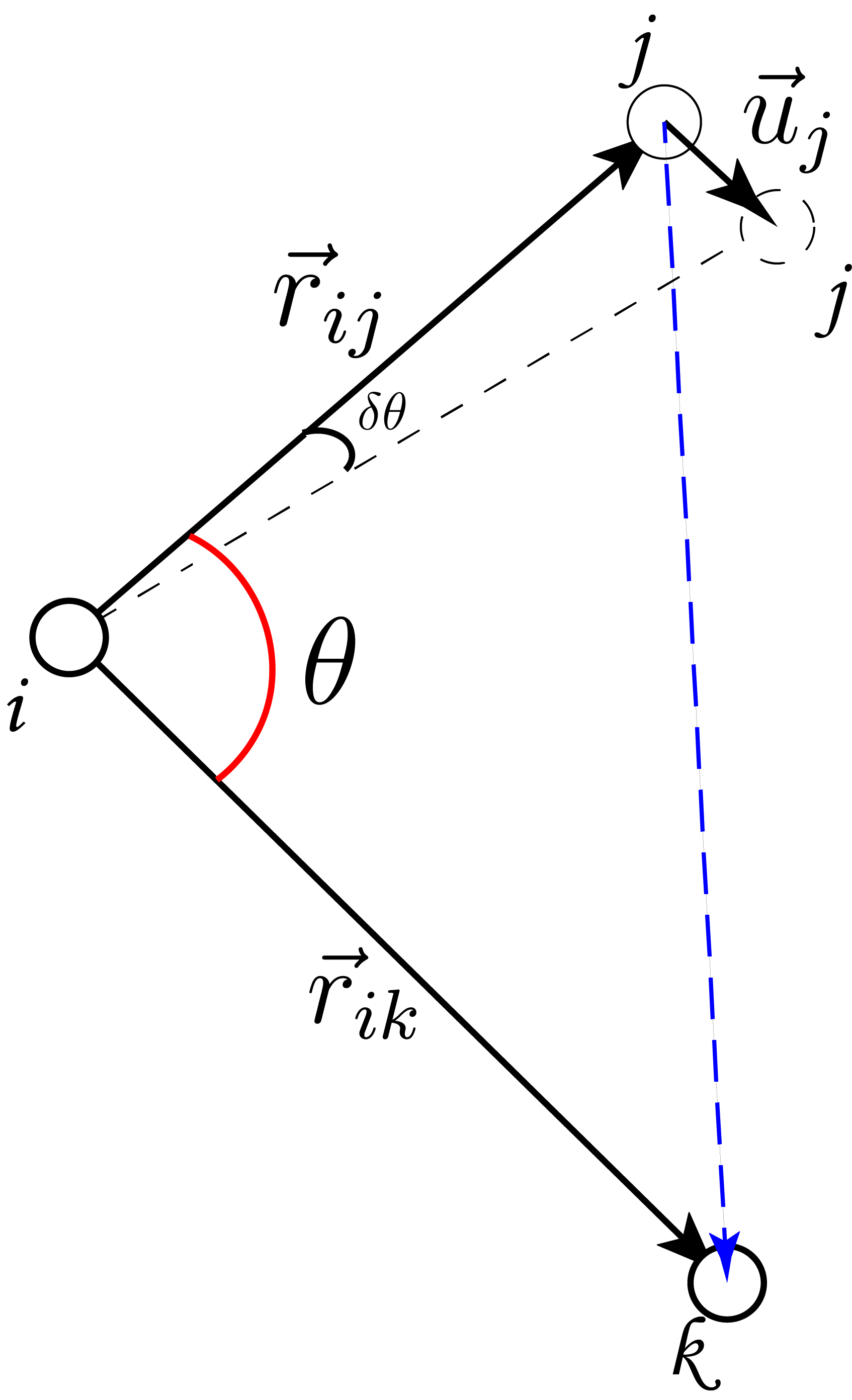}
    \caption{A bond pair in the AS model and its relation to NNN model. Angle $\theta$ (shown in red) is the bond angle between bonds $\mathbf{r}_{ij}$ and $\mathbf{r}_{ik}$.  The  NNN bond $\mathbf{r}_{jk}$ (shown in dashed blue line) is a leading order approximation of the AS. When the site $j$ moves by a small displacement $\mathbf{u}_j$, both the bond angle and the NNN bond length change.  } 
    \label{fig:k_conversion}
\end{figure}

The force derivative $\dv{u}$is taken with respect to the displacements of the sites $\mathbf{u}$ away from their equilibrium positions, and $\alpha$ labels the $x$ and $y$ components of the force. We do not need to concern ourselves with the derivative of the sign function as this only gives a delta function when the angle is $\pi$ due to the singularity.

Note that there is no force if the bond pair is at the rest angle. However, if the bond is bent, there is a force not only on the external particles but also on the central particle, which is only symmetric if the distances between particles are the same.

We can simplify this equation using orthogonal vectors in 2D. For a given vector $\overrightarrow{V}$, the orthogonal vector is provided by 

\begin{equation}
    \overrightarrow{V}^\perp = \left(\begin{matrix}-V_y \\ V_x\end{matrix} \right).
\end{equation}

The force on any particle due to the angular bonds is therefore given by

\begin{equation}
F_{l\alpha}=\sum_{\{i,jk\}} \kappa_{i,jk} \left( \theta_{i,jk}-\theta_{i,jk}^0\right) \left( \frac{(\delta_{jl}-\delta_{il}) \hat{r}_{ij,\alpha}^\perp}{r_{ij}}  -\frac{(\delta_{kl}-\delta_{il}) \hat{r}_{ik,\alpha}^\perp}{r_{ik}} \right).
\end{equation}

\section{$\kappa_\theta$ to $\kappa$ conversion}\label{app:kappa_conversion}

By examining an angle's geometric relations, we can build a conversion relation between the AS spring constant and the NNN spring constant. We do so by examining the conversion between the small angle change $\delta \theta$ of the angle $\theta$ between bonds $\mathbf{r}_{ij}$ and $\mathbf{r}_{ik}$ and the change of length $\delta l$ of bond $\mathbf{r}_{jk}$ between sites $j$ and $k$, which is the NNN springs bond that the AS of angle $\theta$ is approximated as. A diagram of the geometry is shown in Fig.~\ref{fig:k_conversion}.

As the Fig.~\ref{fig:k_conversion} shows, the angle $\theta$ is changed by $\delta \theta$ when the site $j$ goes through a small displacement $\mathbf{u}_j$, however, as the displacement is small, we approximate the length $r_{ij} = \abs{\mathbf{r}_{ij}}$ remains unchanged as the angle is rotated. Then the length change $\delta l$ for the bond length $r_{jk} = \abs{\mathbf{r}_{jk}}$ can be written as 
    \begin{equation} \label{eq:dl}
        \delta l = \sqrt{r^2_{ij} + r^2_{ik} - 2 r_{ij} r_{ik} \cos \qty(\theta)} - \sqrt{r^2_{ij} + r^2_{ik} - 2 r_{ij} r_{ik} \cos \qty(\theta - \delta \theta)}
    \end{equation}
Taking the deformation angle $\delta \theta$ to be small and expand Eq.~\eqref{eq:dl} to linear order in $\delta \theta$, we have
    \begin{equation}
        \delta l \approx \frac{r_{ij} r_{ik} \sin \qty(\theta)}{\sqrt{r^2_{ij} + r^2_{ik} - 2 r_{ij} r_{ik} \cos \qty(\theta)}} \delta \theta,
    \end{equation}
where $r_{ij} = \abs{\mathbf{r}_{ij}}$ and $r_{ik} = \abs{\mathbf{r}_{ik}}$ are the lengths of the two bonds which the angle $\theta$ is composed of. And by this geometric relation between the length change of the NNN spring $\delta l$ and the angle change $\delta \theta$, we conclude with the relation Eq.~\eqref{eq:k_conversion} between the AS constant $\kappa_{\theta}$ and the NNN spring constant $\kappa$ in Sec.~\ref{sec:model}. 

\section{Numerical Methods} \label{app:NMethod}
Two numerical methods were used in the simulations. For the simulations with NN and NNN springs, a simple gradient descent method was used. This implementation calculates the force of the nodes in the system and displaces them following the force gradient until an error tolerance of $10^{-7}$ is reached. As for the simulation including AS, a \emph{Fast Inertial Relaxation Engine (FIRE)} algorithm is used to handle the non-linear multi-particle interactions and to accelerate the computation. For the \emph{FIRE} algorithm, not only the force gradient $\mathbf{F}$ is calculated, but also the velocity $\mathbf{v}$ and its dot product with the force $P = \mathbf{F} \cdot \mathbf{v}$. Then the direction and the speed of the descent are controlled by $P$. 

\end{widetext}

\bibliography{references}

\begin{thebibliography}{10}

\bibitem{liu2000locally}
Liu Z, Zhang X, Mao Y, Zhu Y, Yang Z, Chan CT, et~al.
\newblock Locally resonant sonic materials.
\newblock Science. 2000;289(5485):1734-6.

\bibitem{Huber2016}
Huber SD.
\newblock Topological mechanics.
\newblock Nature Physics. 2016 6;12:621-3.
\newblock Available from: \url{https://www.nature.com/articles/nphys3801}.

\bibitem{Bertoldi2017}
Bertoldi K, Vitelli V, Christensen J, Hecke MV.
\newblock Flexible mechanical metamaterials.
\newblock Nature Reviews Materials. 2017 10;2:1-11.
\newblock Available from:
  \url{https://www.nature.com/articles/natrevmats201766}.

\bibitem{Yu2018}
Yu X, Zhou J, Liang H, Jiang Z, Wu L.
\newblock Mechanical metamaterials associated with stiffness, rigidity and
  compressibility: A brief review.
\newblock Progress in Materials Science. 2018 5;94:114-73.

\bibitem{Xin2020}
Xin L, Siyuan Y, Harry L, Minghui L, Yanfeng C.
\newblock Topological mechanical metamaterials: A brief review.
\newblock Current Opinion in Solid State and Materials Science. 2020
  10;24:100853.

\bibitem{Lakes2017}
Lakes RS.
\newblock Negative-Poisson's-Ratio Materials: Auxetic Solids.
\newblock https://doiorg/101146/annurev-matsci-070616-124118. 2017 7;47:63-81.
\newblock Available from:
  \url{https://www.annualreviews.org/doi/abs/10.1146/annurev-matsci-070616-124118}.

\bibitem{Nicolaou2012}
Nicolaou ZG, Motter AE.
\newblock Mechanical metamaterials with negative compressibility transitions.
\newblock Nature Materials 2012 11:7. 2012 5;11:608-13.
\newblock Available from: \url{https://www.nature.com/articles/nmat3331}.

\bibitem{Kane2013}
Kane CL, Lubensky TC.
\newblock Topological boundary modes in isostatic lattices.
\newblock Nature Physics. 2013 12;10:39-45.
\newblock Available from: \url{https://www.nature.com/articles/nphys2835}.

\bibitem{Lubensky2015}
Lubensky TC, Kane CL, Mao X, Souslov A, Sun K.
\newblock Phonons and elasticity in critically coordinated lattices.
\newblock Reports on Progress in Physics. 2015 6;78:073901.

\bibitem{Mao2018}
Mao X, Lubensky TC.
\newblock Maxwell Lattices and Topological Mechanics.
\newblock Annual Review of Condensed Matter Physics. 2018 3;9:413-33.

\bibitem{Sun2020}
Sun K, Mao X.
\newblock Continuum Theory for Topological Edge Soft Modes.
\newblock Physical Review Letters. 2020 5;124:207601.
\newblock Available from:
  \url{https://journals.aps.org/prl/abstract/10.1103/PhysRevLett.124.207601}.

\bibitem{bilal2017intrinsically}
Bilal OR, S{\"u}sstrunk R, Daraio C, Huber SD.
\newblock Intrinsically polar elastic metamaterials.
\newblock Advanced Materials. 2017;29(26):1700540.

\bibitem{Rocklin2017a}
Rocklin DZ, Zhou S, Sun K, Mao X.
\newblock Transformable topological mechanical metamaterials.
\newblock Nature Communications. 2017 1;8:1-9.
\newblock Available from: \url{https://www.nature.com/articles/ncomms14201}.

\bibitem{Ma2018}
Ma J, Zhou D, Sun K, Mao X, Gonella S.
\newblock Edge Modes and Asymmetric Wave Transport in Topological Lattices:
  Experimental Characterization at Finite Frequencies.
\newblock Physical Review Letters. 2018 8;121:094301.
\newblock Available from:
  \url{https://journals.aps.org/prl/abstract/10.1103/PhysRevLett.121.094301}.

\bibitem{Xiu2022}
Xiu H, Liu H, Poli A, Wan G, Sun K, Arruda EM, et~al.
\newblock Topological transformability and reprogrammability of multistable
  mechanical metamaterials.
\newblock Proceedings of the National Academy of Sciences.
  2022;119(52):e2211725119.

\bibitem{Jolly2022}
Jolly JC, Jin B, Jin L, Lee Y, Xie T, Gonella S, et~al.
\newblock Soft mechanical metamaterials with transformable topology protected
  by stress caching.
\newblock Advanced Science. 2023:2302475.

\bibitem{Bergne2022}
Bergne A, Baardink G, Loukaides EG, Souslov A.
\newblock Scalable 3D printing for topological mechanical metamaterials.
\newblock Extreme Mechanics Letters. 2022 nov;57:101911.
\newblock Available from: \url{https://doi.org/10.1016%2Fj.eml.2022.101911}.

\bibitem{widstrand2023stress}
Widstrand C, Hu C, Mao X, Labuz J, Gonella S.
\newblock Stress focusing and damage protection in topological Maxwell
  metamaterials.
\newblock International Journal of Solids and Structures. 2023;274:112268.

\bibitem{widstrand2023robustness}
Widstrand C, Mao X, Gonella S.
\newblock Robustness of stress focusing in soft lattices under
  topology-switching deformation.
\newblock arXiv preprint arXiv:231014972. 2023.

\bibitem{Paulose2015a}
Paulose J, Meeussen AS, Vitelli V.
\newblock Selective buckling via states of self-stress in topological
  metamaterials.
\newblock Proceedings of the National Academy of Sciences of the United States
  of America. 2015 6;112:7639-44.
\newblock Available from:
  \url{https://www.pnas.org/doi/abs/10.1073/pnas.1502939112}.

\bibitem{Sun2012}
Sun K, Souslov A, Mao X, Lubensky TC.
\newblock Surface phonons, elastic response, and conformal invariance in
  twisted kagome lattices.
\newblock Proceedings of the National Academy of Sciences of the United States
  of America. 2012 7;109:12369-74.
\newblock Available from:
  \url{https://www.pnas.org/doi/abs/10.1073/pnas.1119941109}.

\bibitem{Zhang2018}
Zhang L, Mao X.
\newblock Fracturing of topological Maxwell lattices.
\newblock New Journal of Physics. 2018 6;20:063034.

\bibitem{Zhou2020}
Zhou D, Ma J, Sun K, Gonella S, Mao X.
\newblock Switchable phonon diodes using nonlinear topological Maxwell
  lattices.
\newblock Physical Review B. 2020 3;101:104106.
\newblock Available from:
  \url{https://journals.aps.org/prb/abstract/10.1103/PhysRevB.101.104106}.

\bibitem{Calladine1978}
Calladine CR.
\newblock Buckminster Fuller's "Tensegrity" structures and Clerk Maxwell's
  rules for the construction of stiff frames.
\newblock International Journal of Solids and Structures. 1978 1;14:161-72.

\bibitem{Chen2014}
Chen BGG, Upadhyaya N, Vitelli V.
\newblock Nonlinear conduction via solitons in a topological mechanical
  insulator.
\newblock Proceedings of the National Academy of Sciences of the United States
  of America. 2014 9;111:13004-9.
\newblock Available from:
  \url{https://www.pnas.org/doi/abs/10.1073/pnas.1405969111}.

\bibitem{Guo2019}
Guo XF, Ma L.
\newblock Periodic topological lattice with different indentation hardness on
  opposite surfaces.
\newblock Materials and Design. 2019 10;180:107953.

\bibitem{stenull2019signatures}
Stenull O, Lubensky T.
\newblock Signatures of topological phonons in superisostatic lattices.
\newblock Physical review letters. 2019;122(24):248002.

\bibitem{saremi2020topological}
Saremi A, Rocklin Z.
\newblock Topological elasticity of flexible structures.
\newblock Physical Review X. 2020;10(1):011052.

\bibitem{charara2022omnimodal}
Charara M, McInerney J, Sun K, Mao X, Gonella S.
\newblock Omnimodal topological polarization of bilayer networks: Analysis in
  the Maxwell limit and experiments on a 3D-printed prototype.
\newblock Proceedings of the National Academy of Sciences.
  2022;119(40):e2208051119.

\bibitem{Xiu2023}
Xiu H, Frankel I, Liu H, Qian K, Sarkar S, MacNider B, et~al.
\newblock Synthetically non-Hermitian nonlinear wave-like behavior in a
  topological mechanical metamaterial.
\newblock Proceedings of the National Academy of Sciences.
  2023;120(18):e2217928120.
\newblock Available from:
  \url{https://www.pnas.org/doi/abs/10.1073/pnas.2217928120}.

\bibitem{AltlandZirnbauer}
Altland A, Zirnbauer MR.
\newblock Nonstandard symmetry classes in mesoscopic normal-superconducting
  hybrid structures.
\newblock Phys Rev B. 1997 Jan;55:1142-61.
\newblock Available from:
  \url{https://link.aps.org/doi/10.1103/PhysRevB.55.1142}.

\bibitem{RevModPhys.88.035005}
Chiu CK, Teo JCY, Schnyder AP, Ryu S.
\newblock Classification of topological quantum matter with symmetries.
\newblock Rev Mod Phys. 2016 Aug;88:035005.
\newblock Available from:
  \url{https://link.aps.org/doi/10.1103/RevModPhys.88.035005}.

\end{thebibliography}

\end{document}